\documentclass[aps,prb,twocolumn,preprintnumbers,amsmath,amssymb]{revtex4}
\usepackage[dvips]{graphicx}
\usepackage{dcolumn}

\begin{document}

\title{
Coupled k-space structure of $d$-wave superconducting and magnetic orders induced by paramagnetic pair-breaking effect
}

\author{Yuhki Hatakeyama}
\author{Ryusuke Ikeda}

\affiliation{
Department of Physics, Kyoto University, Kyoto 606-8502, Japan
}

\date{\today}

\begin{abstract}
We theoretically investigate k-space structures of $d_{x^2-y^2}$-wave superconducting (SC) and spin-density-wave (SDW) orders in their coexistent phase induced by a paramagnetic pair-breaking (PPB) effect in relation to the high field and low temperature (HFLT) SC phase in CeCoIn$_5$. 
It is shown that, in k-space, the SDW order develops near the gap nodes where the SC order is suppressed by PPB, and the nesting condition for the SDW ordering is satisfied. 
By comparing the results in the $d_{x^2-y^2}$-wave SC model and those in an artificial model with no sign change of the gap function in k-space with each other, it is shown that the $d_{x^2-y^2}$-wave SC and SDW orders are enhanced altogether in k-space due to the sign change of the $d_{x^2-y^2}$-wave gap function there, and that this mutual enhancement largely stabilizes the coexistence of these orders in real space. 
It is also discussed that the field dependence of a SDW moment can be affected by the k-space structure of these orders, which is dependent on the curvature of the Fermi surface.
\end{abstract}

\maketitle

\section{Introduction}
\label{sec:introduction}

The high field and low temperature (HFLT) superconducting (SC) phase of CeCoIn$_5$ is a novel type of state of matter, and a lot of experimental and theoretical studies have been performed to obtain the genuine picture on this phase. In this phase, the $d_{x^2-y^2}$-wave pairing and a SDW orders coexist, although this SDW order is absent in the normal phase\cite{kenzelmann_2008_kenzelmann_2010} just above $H_{c2}(T)$. 

Early experimental data have already indicated that this SC material has an unusually strong PPB effect. Based on theoretical analysis\cite{adachi_2003,ikeda_2007,ikeda_2010b} of subsequent experimental facts on the vortex lattice rigidity\cite{watanabe_2004}, NMR\cite{kumagai_2011}, and doping effects\cite{tokiwa_2008_tokiwa_2010}, it has been argued that the Fulde-Ferrell-Larkin-Ovchinnikov (FFLO) SC state, where a SC order parameter $\Delta$ is spontaneously modulated in real space due to a strong paramagnetic pair-breaking (PPB)\cite{fulde_1964_larkin_1965}, is realized in the HFLT phase. Further, it has been clarified \cite{ikeda_2010a} that an interplay between the $d_{x^2-y^2}$-wave pairing symmetry and the strong PPB creates a SDW order in contrast to the widespread view that the SC and magnetic orders are competitive with each other. This theoretical picture on the SDW ordering based only on the strong PPB explains the main features of the SDW order observed through a neutron scattering experiment\cite{kenzelmann_2008_kenzelmann_2010,Gerber} and an NMR experiment\cite{kumagai_2011}. 

In our previous works\cite{ikeda_2010a,hatakeyama_2011}, it has been stressed that the PPB-induced coexistence of $d_{x^2-y^2}$-wave SC and the SDW orders occurs, broadly speaking, with two origins. Firstly, suppression of the SDW ordering due to a SC excitation gap is weakened due to PPB, and, secondly, the SDW ordering is enhanced due to the sign change of the $d_{x^2-y^2}$-wave gap function $w_{\bf k}$ in k-space where $w_{\bf k+Q_0}=-w_{\bf k}$ is satisfied with ${\bf Q}_0=(\pi,\pi,\pi)$. 
The importance of k-space sign change of the gap function for the PPB-induced SDW ordering have also been discussed elsewhere\cite{michal_2011} in relation to a PPB-induced shift of a resonance peak. 
Furthermore, it has also been shown that the PPB-induced SDW ordering is largely enhanced by the FFLO modulation of $\Delta$ parallel to the field\cite{hatakeyama_2011}. In addition, it has been shown \cite{hatake2015} that the presence of such a longitudinal FFLO spatial modulation in the HFLT phase explains the switching of the SDW $q$-vector via the in-plane rotation of the applied magnetic field seen in CeCoIn$_5$ \cite{Gerber}. 

Nevertheless, it is surprising that this SDW order in {\it real} space tends to favor the region with a nonvanishing SC order rather than the vicinity of the vortex cores\cite{hatakeyama_2011}. 
Then, one may wonder how the magnetic order favoring coexistence with a non-vanishing SC order in real space can coexist with the SC order even in k-space. 
This naive question has motivated us to investigate details of k-space structure of this PPB-induced SDW order in a $d_{x^2-y^2}$-wave SC phase. 

In this paper, we present a theoretical analysis on k-space structure of the coexisting $d_{x^2-y^2}$-wave SC and SDW orders induced by a strong PPB effect. 
It is found that the SDW order develops in the k-space region near the $d_{x^2-y^2}$-wave gap node, where the SC order is suppressed by PPB, and the nesting condition of the Fermi surface (FS) is satisfied. 
Effects of the sign change of the $d_{x^2-y^2}$-wave gap function around each gap node on the coexistence of these orders are discussed through a comparison with the results of an artificial model in which the gap function with no sign change in k-space is assumed. 
It is shown that, due to the sign change of the gap function, the $d_{x^2-y^2}$-wave SC and SDW orders are enhanced cooperatively even in k-space, and the coexistence of these orders is largely stabilized by this mutual enhancement. 
Moreover, it has been found that, in the case of a large FS curvature, a SDW staggered moment is maximal \textit{slightly below} the SC transition field $H_{c2}$ due to the mismatch of the nesting hotspot and the k-space region where the SC order is suppressed by PPB. 
These results indicate the presence of a profound mechanism of the PPB-induced coexistence of $d_{x^2-y^2}$-wave SC and SDW orders in the light of the k-space structure of these orders. 

The mechanism of the PPB-induced SDW ordering has also been discussed by another research group \cite{kato_2011_kato_2012}, in which the importance of the nesting between quasiparticle pockets appearing in k-space due to strong PPB, on which the lower excitation energy in a pure SC phase is equal to zero, has been stressed. 
The theoretical works of Refs.\onlinecite{ikeda_2010a,hatakeyama_2011,michal_2011,kato_2011_kato_2012} are based on essentially the same model, and thus, if the quasiparticle pockets' contribution is dominant, their contribution might also be reflected in k-space results in the framework of our previous works \cite{ikeda_2010a,hatakeyama_2011}. 
In the sense of clarifying a genuine picture on the coexistence of the two orders, the present comprehensive study on the k-space structure of these coexisting orders is believed to be valuable. 

This paper is organized as follows. 
In Sec.\ref{sec:model}, we explain the formulation of our theoretical model.
In Sec.III, the structure of $d_{x^2-y^2}$-wave SC and SDW orders in k-space is presented. 
In Sec.\ref{sec:sign_change}, influences of the sign change of the gap function on the k-space structure of these orders are discussed, by comparing the results in the $d_{x^2-y^2}$-wave SC model and in the artificial model with one another. 
In Sec.\ref{sec:curvature}, effects of the FS curvature on k-space structure of these orders and its consequences in the field dependence of a SDW staggered moment are discussed. 
In Sec.\ref{sec:summary}, we present a summary and discussions of the results.
Throughout the present paper, expressions are written in unit of $\hbar=k_{\rm B}=c=1$.

\section{Model}
\label{sec:model}

Our starting Hamiltonian including the mean-field interaction channels of SC and SDW orders is expressed as ${\cal H}={\cal H}_{\rm kin}+{\cal H}_{\rm SC}+{\cal H}_{\rm SDW}$, where
\begin{align}
 {\cal H}_{\rm kin}&=\sum_{{\bf k},\sigma}(\xi_{\bf k}-h\sigma)c_{{\bf k},\sigma}^\dagger c_{{\bf k},\sigma} , \label{eq:Hkin} \\
 {\cal H}_{\rm SC}&=\frac{|\Delta|^2}{\lambda}-\left[\Delta\sum_{\bf k} w_{\bf k}c_{{\bf k},\uparrow}^\dagger c_{{\bf -k},\downarrow}^\dagger + {\rm h.c.} \right] \label{eq:HSC} , \\
 {\cal H}_{\rm SDW}&=\frac{m^2}{U}-\left[m\sum_{{\bf k},\sigma} c_{{\bf k},-\sigma}^\dagger c_{{\bf k+Q},\sigma} + {\rm h.c.} \right] \label{eq:HSDW} . 
\end{align}
Here, $c_{{\bf k},\sigma}$ is the annihilation operator of the quasiparticle state labeled by its wave vector ${\bf k}$ and spin projection $\sigma$($=\pm1$), $h=g\mu_{\rm B}H$ is the Zeeman energy ($g$ is the g-factor, $\mu_{\rm B}$ is the Bohr magneton, and $H$ is the strength of the magnetic field directed perpendicularly to the c-axis, $\Delta$ is the SC order parameter, $\lambda$ is the strength of the SC pairing interaction, $w_{\bf k}$ is the gap function, $m$ is the SDW staggered moment, $U$ is the strength of the exchange interaction, and ${\bf Q}$ is the SDW modulation wave vector. 
The dispersion relation $\xi_{\bf k}$ is that of the two-dimensional lattice system and is given by
\begin{equation}
 \xi_{\bf k}=-2t_1(\cos(k_x)+\cos(k_y))-4t_2\cos(k_x)\cos(k_y)-\mu ,
\end{equation}
where $t_1$ and $t_2$ are the nearest and next-nearest neighbor hopping, respectively, and $\mu$ is the chemical potential, which is adjusted so that the FS nesting condition $\xi_{\bf k+Q}=-\xi_{\bf k}$ is precisely satisfied at the $d_{x^2-y^2}$-wave gap nodes on the line $k_x=k_y$ if ${\bf Q}=(\pi, \pi)$ (see, however, below). 
Our calculation is performed in the Pauli limit, i.e., the PPB effect is assumed to be so strong that the orbital pair-breaking effect is negligible.
In accord with a neutron scattering experiment in the HFLT phase of CeCoIn$_5$\cite{kenzelmann_2008_kenzelmann_2010}, it is assumed that the SDW staggered moment is parallel to the c-axis, and that a single-${\bf Q}$ SDW order is always realized \cite{hatake2015}. 

To stress that the PPB-induced SDW ordering in the $d$-wave SC phase does not stem from a specific Fermi surface but is a general phenomenon insensitive to the details of the Fermi surface, the simplest Fermi surface of the 2D tight-binding model will be used. Further, a FFLO spatial modulation of the SC order parameter, believed to be present in the high field phase of CeCoIn$_5$, will not be considered, and the SDW ordering in this work will be examined by assuming a spatially homogeneous SC order parameter because a SDW or antiferromagnetic ordering enhanced in higher fields is not believed to be peculiar to CeCoIn$_5$: As argued elsewhere \cite{ikeda_2010a,hatakeyama_2011}, the tendency of the SDW ordering at the high field end of the SC phase is also seen in heavy fermion superconductors other than CeCoIn$_5$. 
 
Regarding the form of the pairing function $w_{\bf k}$, the following two forms will be considered in the present paper: the first one is the usual $d_{x^2-y^2}$-wave one $w_{\bf k}=\cos(k_x)-\cos(k_y)$, which changes its sign in k-space such that $w_{{\bf k+Q}_0}=-w_{\bf k}$ with ${\bf Q}_0=(\pi,\pi)$, and the second one is of an extended $s$-wave type $w_{\bf k}=|\cos(k_x)-\cos(k_y)|$, which does not change its sign in k-space. 
The implication of the use of the second functional form will be explained later in Sec.\ref{sec:sign_change}.

In our calculation, it is assumed that the SDW order is incommensurate so that  the modulation vector ${\bf Q}$ is equal to $\pi(1+1/N,1+1/N)$, where $N$ is an integer. Note that, due to the four-fold rotational symmetry of $\xi_{\bf k}$ and $|w_{\bf k}|$, the $\pi/2$-rotation of ${\bf k}$ transforms ${\cal H}$ with ${\bf Q}=\pi(1 + 1/N, -1 - 1/N)$ to ${\cal H}$ with ${\bf Q}=\pi(1 + 1/N, 1 + 1/N)$. Therefore, it is justified to restrict ourselves to the case with ${\bf Q}=\pi(1 + 1/N, 1 + 1/N)$. 
With this value of ${\bf Q}$, $4N$ quasiparticle states are coupled with one another in ${\cal H}_{\rm SC}$ and ${\cal H}_{\rm SDW}$, and thus the action ${\cal S}$ corresponding to the Hamiltonian ${\cal H}$ is written as a bilinear form of $4N$-dimension
\begin{equation} 
	{\cal S}=\frac{T}{2N}\sum_{k} \hat{\Psi}_k^\dagger \hat{G}^{-1}_{k} \hat{\Psi}_k+\frac{1}{T} \left(\frac{|\Delta|^2}{\lambda}+\frac{m^2}{U} \right),
	\label{eq:action}
\end{equation}
where
\begin{align}
	\hat{\Psi}_k =\big( & c_{k,\uparrow}^\dagger , c_{-k,\downarrow} , c_{k+Q,\downarrow}^\dagger , 
	c_{-k-Q,\uparrow} , 
	\nonumber \\ & \quad 
	c_{k+2Q,\uparrow}^\dagger , c_{-k-2Q,\downarrow} ,
	 \ldots , c_{-k-(2N-1)Q,\uparrow} \big) ,
\end{align}
\begin{equation}
	\hat{G}^{-1}_k =
	\begin{pmatrix}
		\hat{D}_{k,\uparrow} & \hat{M} & 0 &   \ldots &  \hat{M} \\
		\hat{M} & \hat{D}_{k+Q,\downarrow} & \hat{M}  &  \ldots & 0 \\
		0 & \hat{M} & \hat{D}_{k+2Q,\uparrow} & \ldots & 0\\
		\vdots & \vdots & \vdots & \ddots & \vdots \\
		\hat{M} & 0 & 0 & \ldots & \hat{D}_{k+(2N-1)Q,\downarrow} 
	\end{pmatrix} , 
\label{matrixG}
\end{equation}
with $k=(i\omega_n,{\bf k})$ ($\omega_n=(2n+1)\pi T$ is the fermion Matsubara frequency), $\hat{D}_{k,\sigma} =(i\omega_n+h\sigma)\hat{I}-\xi_{\bf k}\hat{\sigma}_z+\Delta w_{\bf k}\sigma\hat{\sigma}_x$ and $\hat{M} =m\hat{\sigma}_z$ ($\hat{I}$ is the  $2\times2$ identity matrix, and $\hat{\sigma_i}$ ($i=x,y,z$) is the Pauli matrix).
Then, the free energy density ${\cal F}$ can be calculated straightforwardly based on ${\cal S}$, as
\begin{equation}
	{\cal F}(\Delta,m)=\frac{|\Delta|^2}{\lambda}+\frac{m^2}{U}-\frac{T}{2N}\sum_k \ln \left| \hat{G}_k^{-1} \right| . 
	\label{eq:free_energy}
\end{equation}
In the numerical calculation, we used $N=10$, i.e., ${\bf Q}=2\pi(0.55,0.55)$ in accord with the observed SDW modulation vector in CeCoIn$_5$\cite{kenzelmann_2008_kenzelmann_2010}, and the values of $\Delta$ and $m$ are obtained by numerically minimizing ${\cal F}(\Delta,m)$.

Structures of coexisting SC and SDW orders in k-space can be investigated in terms of the following correlation functions associated with the quasiparticle states with wave vector ${\bf k}$:
\begin{align}
	H({\bf k})&=-\sum_{\sigma} \left[ \langle c_{{\bf k},-\sigma} c^{\dagger}_{{\bf k+Q},\sigma} \rangle+\langle c_{{\bf k},-\sigma} c^{\dagger}_{{\bf k-Q},\sigma} \rangle \right]
	\nonumber \\
	&=T\sum_{\omega_n} \left( [\hat{G}_k+\hat{G}_{k-Q}^*]_{3,1}-[\hat{G}_{-k}+\hat{G}_{-k+Q}^*]_{4,2} \right) ,
	\\
	F({\bf k})&=-\langle c_{{\bf -k},\downarrow} c_{{\bf k},\uparrow} \rangle
	=T\sum_{\omega_n} [\hat{G}_k]_{2,1} ,
\end{align}
where $\hat{G}_k$ is the inverse matrix of $\hat{G}^{-1}_k$.
In a normal phase with no SC and SDW orders, $H({\bf k})$ and $F({\bf k})$ are always zero. 
In an ordered phase, on the other hand, these correlation functions can be finite in the presence of the mean-field couplings (See eqs.(2) and (3)), and the strength of the ordering in k-space can be measured by these correlation functions.
In fact, the gap equations for SC and SDW orders are expressed in terms of $H({\bf k})$ and $F({\bf k})$ as
\begin{equation}
 	\frac{m({\bf r}=0)}{U}=\sum_{\bf k} H({\bf k}) ,
 	\label{eq:m}
 \end{equation}
 \begin{equation}
 	\frac{\Delta}{\lambda}=\sum_{\bf k} w_{\bf k}F({\bf k}) .
 	\label{eq:Delta}
 \end{equation}
Equation (\ref{eq:m}) shows that the SDW staggered moment $m$ is proportional to the sum of $H({\bf k})$ over k-space, indicating that $H({\bf k})$ measures the contribution of the quasiparticle states with wave vector ${\bf k}$ to the SDW ordering. 
In our calculation, we verified that SDW correlation functions with higher harmonics (e.g. $\langle c_{{\bf k},-\sigma} c_{{\bf k+3Q},\sigma}^\dagger \rangle$) are negligibly small, and thus, we focused only on $H({\bf k})$ to study SDW ordering in k-space.
Similarly, Eq.(\ref{eq:Delta}) shows that the SC order parameter $\Delta$ is proportional to the weighted sum of $F({\bf k})$ with the gap function $w_{\bf k}$ over k-space, indicating that $F({\bf k})$ measures the contribution of the quasiparticle states with wave vector ${\bf k}$ to the Cooper pairing.

\section{k-space structure of coexisting SC and SDW orders induced by PPB} 


\begin{figure}[tb]
\scalebox{0.85}[0.85]{\includegraphics[width=0.8\linewidth]{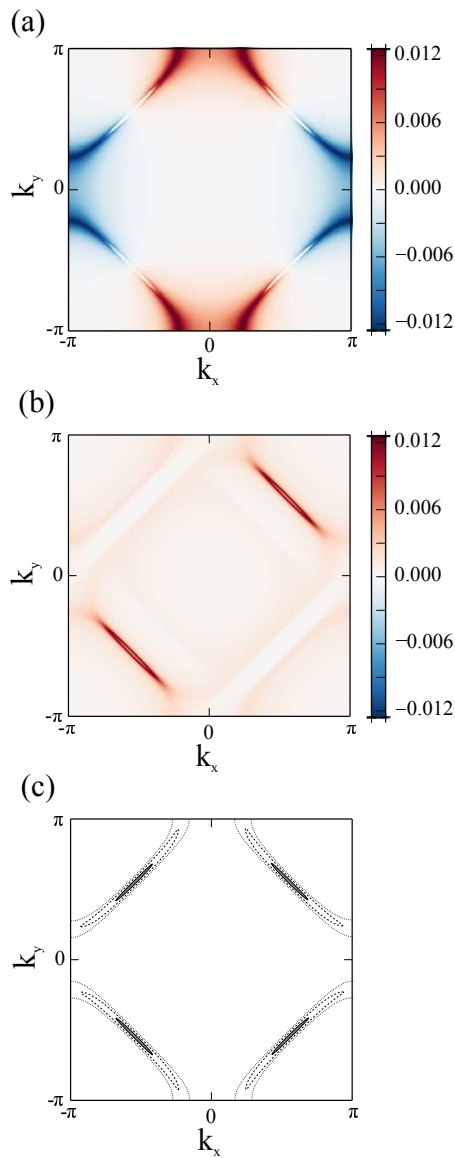} }
\caption{
Colormap plots of (a)$F({\bf k})$, (b)$H({\bf k})$, and (c)the contour lines of $E_{{\bf k},\uparrow}$ in the PPB-induced coexistent phase of $d_{x^2-y^2}$-wave SC and SDW orders. In (c), the isoenergy curves for $E_{{\bf k},\uparrow}/T_c=0$, $1.0$, and $2.0$ are shown as the solid, dashed, and dotted curve, respectively. Inside the solid curve (gray-colored region), $E_{{\bf k},\uparrow} < 0$. 
The used parameters are $t_1/T_c=10$, $t_2/t_1=0.05$, $\mu/t_1=0.62$, $U/T_c=15.2$, $T/T_c=0.1$, and $H/H_{\rm P}=0.85$, where $H_{\rm P}=2.5T_c/g\mu_{\rm B}$ is the Pauli limiting field, and $T_c$ is the SC transition temperature at 
$H=0$.
}. 
\label{fig:map}
\end{figure}

In this section, we present k-space structure of coexisting SC and SDW orders induced by the PPB effect. Examples of k-space distributions of $F({\bf k})$ and $H({\bf k})$ are shown in Fig.1 (a) and (b), respectively, as colormaps. 
Since the major contributions to SC and SDW orderings come from quasiparticle states in the vicinity of FS, the distributions of $F({\bf k})$ and $H({\bf k})$ are concentrated near FS in Fig.\ref{fig:map}.
In Fig.\ref{fig:map}(a), the magnitude of $F({\bf k})$ is strongly suppressed in the narrow oval regions near the $d_{x^2-y^2}$-wave gap nodes (${\bf k}\sim \pm(\pi/2, \pi/2),\pm(\pi/2, -\pi/2)$). 
In these regions, the excitation energy in a pure SC phase with the Zeeman energy 
\begin{equation} E_{{\bf k},\sigma}=\sqrt{\xi_{\bf k}^2+|\Delta w_{\bf k}|^2}-h\sigma \label{eq:Ek}
\end{equation}
become negative for $\sigma=\uparrow$ in higher fields.
Since the negative excitation energy $E_{{\bf k},\uparrow}<0$ means that the Cooper pairing including the quasiparticle with $({\bf k},\uparrow)$ is unstable, $F({\bf k})$ is strongly suppressed in these regions.
However, while $F({\bf k})$ in a pure SC phase is completely suppressed in these regions, $F({\bf k})$ is \textit{not completely} suppressed in these regions when a SDW order is present (see Fig.\ref{fig:FH}(a) and the discussions in Sec.\ref{sec:sign_change}).

In Fig.\ref{fig:map}(b), the major contributions to $H({\bf k})$ come from the narrow oval regions near ${\bf k}\sim\pm(\pi/2, \pi/2)$.
These regions contributing to the SDW ordering are determined under the following two conditions. 
The first one is that the FS nesting condition ($\xi_{\bf k+Q}\sim -\xi_{\bf k}$) is satisfied better in these regions. 
In Fig.\ref{fig:map}, in which the FS curvature is small, this condition is satisfied in a larger area of the FS in the first and third quadrants of the k-space. In relation to this, the case with a large FS curvature will be discussed in 
Sec.\ref{sec:curvature} later. 
The second one is that the SC order is suppressed by PPB in these regions: As the field is increased, the k-space regions contributing to the SDW ordering become broader, while the SC order is suppressed by PPB there. 
These conditions suggest that the SDW ordering is induced mainly by quasiparticle states in the k-space regions where the SC order is suppressed by PPB and the FS nesting condition is satisfied better. In nonzero fields where PPB is not negligible, however, these conditions does not play quantitatively main roles, and rather, as will be discussed in the next section, the $k$-dependent sign change of the $d_{x^2-y^2}$-wave gap function becomes an indispensable condition for stabilizing coexistence of SC and SDW orders in k-space.

\section{Effect of k-space sign change of the gap function}
\label{sec:sign_change}

In this section, we explain the roles of the k-dependent sign change of the $d_{x^2-y^2}$-wave gap function ($w_{\bf k}=\cos(k_x)-\cos(k_y)$) in the PPB-induced coexistence of SC and SDW orders. 
For this purpose, we also analyzed a toy model in which the gap function with no sign change in k-space ($w_{\bf k}=|\cos k_x-\cos k_y|$) is assumed. 
This form of $w_{\bf k}$ is nonanalytic in ${\bf k}$ and might be artificial as a theoretical expression of a gap function of an existing real material. On the other hand, this form qualitatively describes an extended $s$-wave pairing case, and, by comparing the results from this $w_{\bf k}$-form with those from the $d_{x^2-y^2}$-wave pairing form, roles of the amplitude and phase components of $w_{\bf k}$ can be separated from each other to examine the essential roles of the sign change of $w_{\bf k}$ around a gap node in k-space for the SDW ordering. In relation to this, note that the SC quasiparticle excitation energy $E_{{\bf k},\sigma}$ is insensitive to the sign of $w_{\bf k}$ so that $E_{{\bf k},\sigma}$ in the artificial pairing-model mentioned above is identical to that in the usual $d_{x^2-y^2}$-wave case (see Eq.(\ref{eq:Ek})). Consequently, the k-space region where the SC order is suppressed by PPB ($E_{{\bf k},\uparrow}<0$) in the artificial model becomes completely the same as that in the $d_{x^2-y^2}$-wave case. As is seen below, the SDW ordering is not sensitive much to the ${\bf k}$-dependence of $E_{{\bf k},\sigma}$. 

\begin{figure}[tb]
\scalebox{0.23}[0.23]{\includegraphics{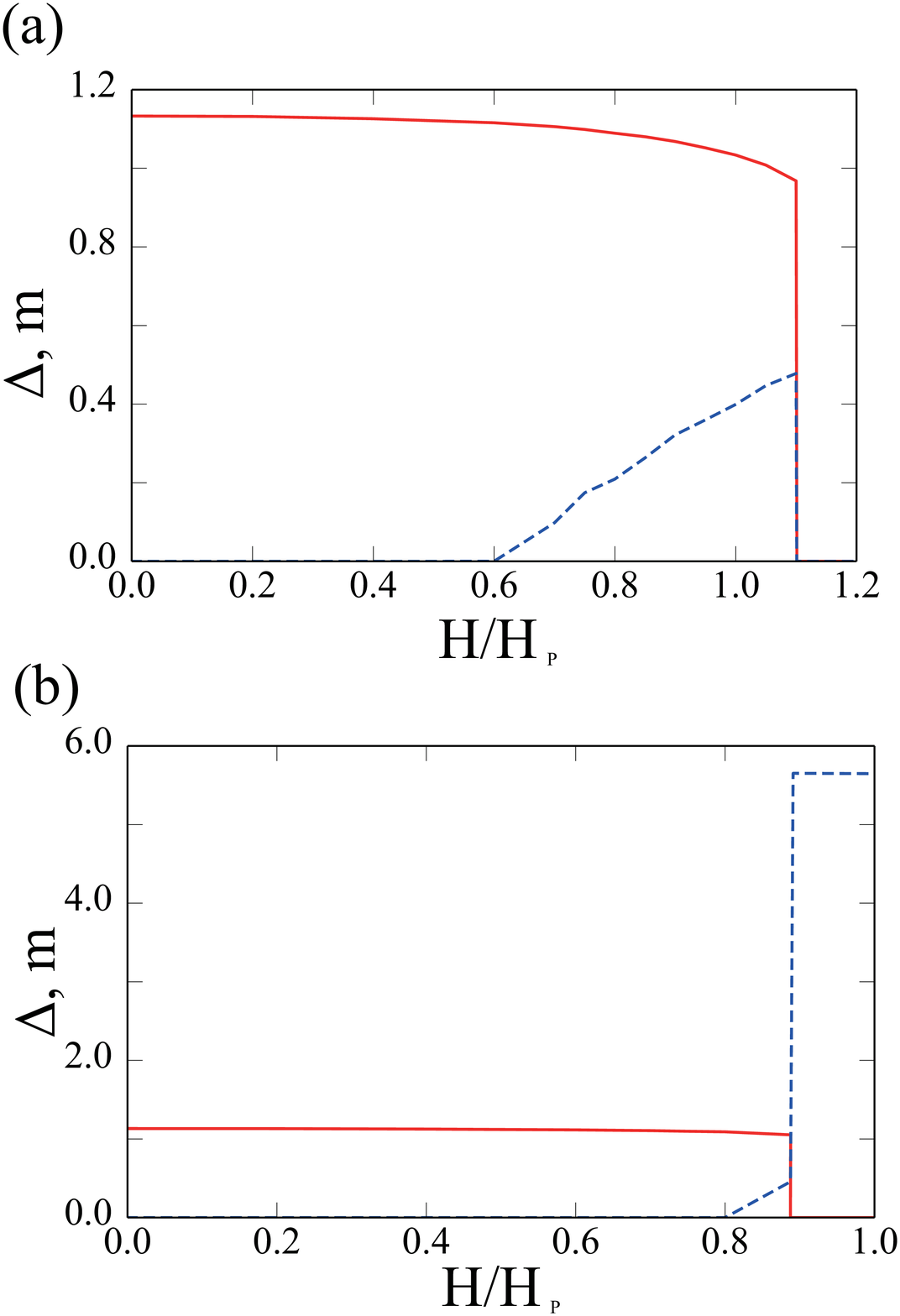} }
\caption{
Field dependences of $\Delta$ (red solid line) and $m$ (blue dashed line) (a) in the $d_{x^2-y^2}$-wave SC model and (b) in the artificial model (see the main text). 
The used parameters in (a) are $t_1/T_c=10$, $t_2/t_1=0.05$, $\mu/t_1=0.62$, $U/T_c=15.2$, and $T/T_c=0.1$. The used parameters in (b) are the same as those in (a) except for $U/T_c=18.8$
}
\label{fig:OPvsH}
\end{figure}

Figure \ref{fig:OPvsH} shows the field dependences of $\Delta$ and $m$ in the $d_{x^2-y^2}$-wave SC model (Fig.\ref{fig:OPvsH}(a)) and in the artificial model (Fig.\ref{fig:OPvsH}(b)).
In order to make the conditions, other than the gap function $w_{\bf k}$, equal between both the models , we used the common parameter values regarding the electronic details in these models. 
In spite of this, a large difference is found in the field dependences of $m$ between these models. 
In Fig.\ref{fig:OPvsH}(a), the SDW ordering is not realized in the high-field normal (non SC) state, and the coexistence phase of SC and SDW orders is induced by PPB in the high-field region of the SC phase. 
In Fig.\ref{fig:OPvsH}(b), on the other hand, the SDW ordering is strongly suppressed in the SC phase compared to in the high-field normal phase, which indicates a clear competition between SC and SDW orders in the artificial model.
This difference implies that the sign change of the gap function, which is present in the $d_{x^2-y^2}$-wave SC model and not in the artificial model, is essential to the realization of the PPB-induced SDW ordering only in a SC phase.

In both figures of Fig.2, the appearance of the nonvanishing $m$ with increasing $h$ is continuous. This second order transition to the SDW ordered phase might be seen as a kink of the field dependence of $\Delta$. Since the SC order is already present at this second order transition field where $\Delta$ is large, however, effects of $m$ on $\Delta$ are much smaller contributions to the free energy compared with those of $\Delta$ on $m$ at least for the parameter values used in obtaining Fig.2. Hence, such a kink is not visible on the field scale in Fig.2. 

\begin{figure}[tb]
\scalebox{0.23}[0.23]{\includegraphics{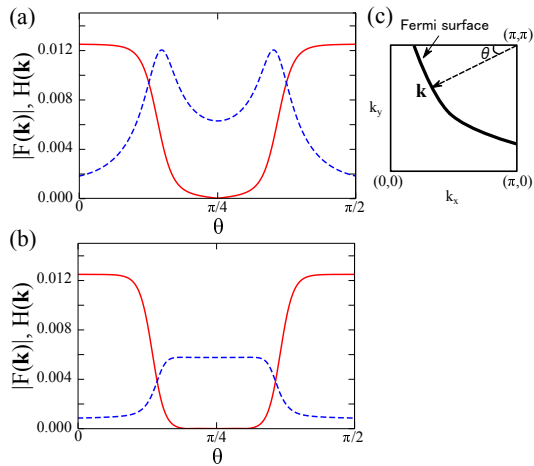} }
\caption{
Plots of $|F({\bf k})|$ (red solid line) and $H({\bf k})$ (blue dashed line) along FS (a) in the $d_{x^2-y^2}$-wave SC model and (b) in the artificial model (see main text) when $H/H_{\rm P}=0.85$. The definition of $\theta$ is illustrated in (c). The used parameters in (a) and (b) are the same as those in Fig.\ref{fig:OPvsH} (a) and (b), respectively.
}
\label{fig:FH}
\end{figure}

The effect of the sign change of the gap function is also reflected in the k-space structures of SC and SDW orders. 
Figure \ref{fig:FH} shows the distributions of $|F({\bf k})|$ and $H({\bf k})$ along the FS in the first quadrant of k-space, in the $d_{x^2-y^2}$-wave SC model (Fig.\ref{fig:FH}(a)) and in the artificial model (Fig.\ref{fig:FH}(b)) at $H/H_{\rm P}=0.85$.
In these figures, $\theta$ represents the angle of ${\bf k}$ on the FS measured from the point ($\pi$,$\pi$), as illustrated in Fig.\ref{fig:FH}(c).
In Fig.\ref{fig:FH}(a), $H({\bf k})$ is maximal not at the gap node (at $\theta=\pi/4$) but at the points away from the gap node where $F({\bf k})$ is finite. 
Moreover, $F({\bf k})$ near the gap node, which is completely suppressed by PPB when there is no SDW order, is slightly {\it enhanced} in the presence of nonvanishing $H({\bf k})$. 
These results imply that SC and SDW orders are enhanced consistently with each other not only in real space but also 
in k-space in the $d_{x^2-y^2}$-wave SC model. 
In Fig.\ref{fig:FH}(b), on the other hand, $H({\bf k})$ is maximal at the gap node, and is strongly suppressed in the region away from the gap node where $F({\bf k})$ is finite. 
Furthermore, $F({\bf k})$ near the gap node is completely suppressed even in the presence of a SDW order. 
These results suggest that, contrary to the $d_{x^2-y^2}$-wave SC model, SC and SDW orders are competitive in k-space in the artificial model. 

The difference in the field dependences of $m$ between these models can be understood in the light of the difference in the k-space structure of SC and SDW orders between these models. 
In the $d_{x^2-y^2}$-wave SC model with a strong PPB effect, SC and SDW orders are enhanced with each other in k-space, and the free energy is lowered by the overlap of these orders in k-space. 
As a result, even if there is no SDW ordering in a normal phase, a SDW order is induced in higher fields by the presence of a SC order.
In the artificial model, on the other hand, SC and SDW orders are competitive in k-space; therefore, a phase with only one kind of orders tends to be more stable than a phase where both SC and SDW orders are present and coexist in k-space.
As a result, the SDW ordering in a SC phase is strongly suppressed in contrast to that in a normal phase. 
As discussed above, the difference between these models is solely originated from the presence or absence of the k-space sign change of the gap function.
Consequently, these results indicate that the k-space sign change of the gap function $w_{\bf k+Q}=-w_{\bf k}$ is the main origin of the mutual enhancement of $d_{x^2-y^2}$-wave SC and SDW orders in k-space and is necessary to realization of the SDW ordering appearing only in a SC phase.

\section{Coexistence of SC and SDW orders close to a continuous SDW transition line}

In this section, the coexistence of the SC and SDW orders in k-space is investigated analytically. For this purpose, we focus here on the case with a small enough $m$ without treating eq.(\ref{eq:action}) in a general way, where $m$ is the amplitude of the SDW order parameter. Namely, the case in which the SDW transition in the SC phase is continuous will be considered in this section. For such a small enough $m$, ${\hat G}^{-1}_k$ defined in eq.(\ref{matrixG}) is simplified to 
\begin{equation}
	\hat{G}_k = \left([ \hat{G}^{(0)}_k ]^{-1} + \check{M} \right)^{-1}
	\sim \hat{G}^{(0)}_k - \hat{G}^{(0)}_k\check{M}\hat{G}^{(0)}_k, 
	\label{eq:Gm}
\end{equation}
where 
\begin{equation}
	[ \hat{G}^{(0)}_k ]^{-1} =
	\begin{pmatrix}
		\hat{D}_{k,\uparrow} & 0 &   \ldots &  0 \\
		0 & \hat{D}_{k+Q,\downarrow} &  \ldots  & 0 \\
		\vdots & \vdots   & \ddots  & \vdots\\
		0 & 0 &\ldots & \hat{D}_{k+(N-1)Q,\downarrow}, 
	\end{pmatrix} \\	
\end{equation}
and 
\begin{equation}
	\check{M} =
	\begin{pmatrix}
		0 & \hat{M} & 0 &  \ldots & 0 &\hat{M} \\
		\hat{M} & 0 & \hat{M}  & \ldots &0 & 0 \\
		\vdots & \vdots & \vdots  & \ddots & \vdots & \vdots\\
		\hat{M} & 0 & 0 &\ldots & \hat{M} & 0. 
	\end{pmatrix} \\
\end{equation}

Equation (\ref{eq:Gm}) can be regarded as the perturbative expansion of ${\hat G}_k$ in $m$ based on the use of the Nambu Green's function $\hat{D}_{k,\uparrow}^{-1}$ (see below). According to eq.(\ref{eq:Gm}), the components of $\hat{G}_k$ associated with $H({\bf k})$ take the following forms 
\begin{equation}
	[ \hat{G}_k ]_{3,1}
	= [ \hat{D}_{k+Q,\downarrow}^{-1}\hat{M}\hat{D}_{k,\uparrow}^{-1} ]_{1,1}
	= m \left(
	G^{(0)}_{k,\uparrow} G^{(0)}_{k+Q,\downarrow} 
- F^{(0)}_{k,\uparrow} F^{(0)}_{k+Q,\downarrow} \right), 
	\label{eq:G31}
\end{equation}
and 
\begin{equation}
	[ \hat{G}_k ]_{4,2}
	= [ \hat{D}_{k+Q,\downarrow}^{-1}\hat{M}\hat{D}_{k,\uparrow}^{-1} ]_{2,2}
	= m \left( F^{(0)}_{k,\uparrow} F^{(0)}_{k+Q,\downarrow}
	-G^{(0)}_{-k,\downarrow}G^{(0)}_{-k-Q,\uparrow} \right), 
	\label{eq:G42}
\end{equation}
where 
\begin{widetext}
\begin{equation}
	\begin{pmatrix}
		G^{(0)}_{k,\sigma} & F^{(0)}_{k,\sigma} \\
		F^{(0)}_{k,\sigma} & -G^{(0)}_{-k,-\sigma} \\
	\end{pmatrix}
	= \hat{D}_{k,\sigma}^{-1} 
	=\frac{1}{(i\omega_n+h\sigma)^2-\xi_{\bf k}^2-|\Delta w_{\bf k}|^2} 
	\begin{pmatrix}
		i\omega_n+h\sigma+\xi_{\bf k}  & -\Delta w_{\bf k} \sigma \\
		-\Delta w_{\bf k} \sigma & i\omega_n+h\sigma-\xi_{\bf k} \\
	\end{pmatrix}
	\label{eq:G0}
\end{equation}
\end{widetext}
are the Nambu Green's functions with $m=0$. By substituting eqs.(\ref{eq:G31}) and (\ref{eq:G42}) into eq.(9), $H({\bf k})$ is obtained in the form
\begin{equation}
	H({\bf k})
	=H^{(n)}({\bf k}) + H^{(an)}({\bf k}), 
	\label{eq:Hm}
\end{equation}
where 
\begin{equation}
	H^{(n)}({\bf k})=T\sum_{\omega_n,\sigma} 
	m G^{(0)}_{k,\sigma}G^{(0)}_{k+Q,-\sigma}
	+(Q \leftrightarrow -Q), 
	\label{eq:Hm_n}
\end{equation}
and 
\begin{equation}
	H^{(an)}({\bf k})=-T\sum_{\omega_n,\sigma} 
	m F^{(0)}_{k,\sigma}F^{(0)}_{k+Q,-\sigma}
	+(Q \leftrightarrow -Q) 
	\label{eq:Hm_an}
\end{equation}
are contributions from the normal and anomalous Green's functions, respectively. Further, by substituting eq.(\ref{eq:Hm}) into eq.(11), we obtain 
\begin{widetext}
\begin{equation}
	\frac{1}{U}=
	T\sum_{k,\sigma} 
	\frac{\omega_n^2-h^2+\xi_{\bf k}\xi_{\bf {k+Q}}-|\Delta|^2 w_{\bf k} w_{\bf {k+Q}}}
	{((i\omega_n+h\sigma)^2-\xi_{\bf k}^2-|\Delta w_{\bf k}|)
		((i\omega_n-h\sigma)^2-\xi_{\bf {k+Q}}^2
-|\Delta w_{\bf {k+Q}}|)}. 
	\label{eq:gapm}
\end{equation}
\end{widetext}
The normal part $H^{(n)}({\bf k})$ is significantly enhanced in the vicinity of the k-space regions where $E_{{\bf k},\uparrow}$ and $E_{{\bf {k+Q}},\uparrow}$ tend to vanish. This condition is satisfied close to the nodal region where ${\bf k}=\pm(\pi/2,\pi/2)$ and is compatible with the fact shown in $\S 3$ that the SDW order is induced in these k-space regions. 
On the other hand, $H^{(an)}({\bf k})$ expresses effects of the coupling between the Cooper pair condensates and the SDW order and, according to eqs.(\ref{eq:Hm_an}) and (\ref{eq:G0}), is proportional to $-|\Delta|^2 
w_{\bf k} w_{\bf {k+Q}}$. Then, in the case of $d_{x^2-y^2}$-wave pairing satisfying $w_{\bf {k+Q}}=-w_{\bf k}$, the sign of this term becomes positive so that $H({\bf k})$ is enhanced. This $H({\bf k})$-enhancement is known to be an origin of coexistence of the SDW order with the $d_{x^2-y^2}$-wave SC order \cite{KatoMachida}. In fact, it is easily verified that eq.(\ref{eq:gapm}) with $h=0$ and ${\bf Q}\rightarrow(\pi,\pi)$ coincides with the gap equation derived in Ref.23 in $m \to 0$ limit. 

In zero field ($h=0$) studied previously \cite{KatoMachida}, however, this SDW-enhancement close to the gap nodes is much weaker than the SDW-suppression due to the finite energy gap far from the gap nodes, and thus, the SDW ordering {\it only} in the SC phase seen in CeCoIn$_5$ is rarely realized. On the other hand, in high fields of the order of the Pauli-limiting field $H_{\rm P}$, the SDW-enhancement close to the gap nodes becomes more dominant, and the SDW order only in the SC phase can occur. 

By contrast, in the pairing model $w_{\bf k}=|{\rm cos}(k_x) - {\rm cos}(k_y)|$ simulating an extended $s$-wave pairing function, even the contributions from the vicinity of the gap nodes also suppress the SDW ordering, and the SC and SDW orders become competitive in any region of the real and k-spaces. The remarkable differences in the field dependence of $m$ between Figs.2(a) and 2(b) (see also Fig.4(b)) are explained based on the coupling between the two orders dependent on the pairing symmetry and on its dependence on the electron correlation strength $U$: For the relatively small $U$ values used in Fig.2(a), the SDW ordering does not occur in the normal phase, although the SDW order occurs at the high field end of the SC phase due to the SDW-enhancement of the $d_{x^2-y^2}$-wave pairing mentioned above.

\section{Influence of a FS curvature on SDW ordering}
\label{sec:curvature}

\begin{figure}[tb]
\scalebox{0.25}[0.25]{\includegraphics{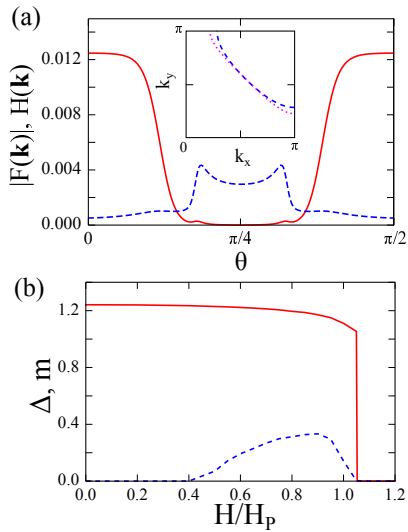} }
\caption{
(a) Plots of $|F({\bf k})|$ (red solid line) and $H({\bf k})$ (blue dashed line) along the FS when $t_2/t_1=-0.05$, $H/H_{\rm P}=1.0$ and $T/T_c=0.1$. 
The definition of $\theta$ is the same as in Fig. \ref{fig:FH}.
Inset: Shapes of FSs in the first quadrant of the k-space for $t_2/t_1=-0.05$ (blue dashed line) and $t_2/t_1=0.05$ (purple dotted line).
(b) Field dependence of the SC order parameter $\Delta$ (red solid line) and the SDW staggered moment $m$ (blue dashed line) when $t_2/t_1=-0.05$ and $T/T_c=0.1$. 
The used parameters are $t_1/T_c=10$, $\mu/t_1=0.63$, and $U/T_c=21.1$ in both (a) and (b). 
}
\label{fig:t2FS}
\end{figure}

In this section, we discuss an influence of a FS curvature on the k-space structure of $d_{x^2-y^2}$-wave SC and SDW orders. 
Since the SC and SDW orders in k-space are induced by quasiparticle states near FS (see Fig.\ref{fig:map}), it is expected that the k-space structure of these orders is affected by the shape of FS. 
In the model described in Sec.\ref{sec:model}, the shape of FS is parametrized by $t_2/t_1$, because $\mu$ is determined under the condition that the FS nesting is realized at the gap nodes on the line $k_x=k_y$.
The inset of Fig.\ref{fig:t2FS}(a) plots the shapes of FSs for $t_2/t_1=-0.05$ (blue dashed line) and $t_2/t_1=0.05$ (purple dotted line), showing that the FS curvature for $t_2/t_1=-0.05$ is larger than that for $t_2/t_1=0.05$.

Figure \ref{fig:t2FS}(a) shows $|F({\bf k})|$ and $H({\bf k})$ along the FS in the case of the large FS curvature ($t_2/t_1=-0.05$) when $H/H_{\rm P}=1.0$.
As discussed in Sec.III, in the case of a small FS curvature, the FS nesting condition ($\xi_{\bf k+Q}\sim -\xi_{\bf k}$) is satisfied in a large part of the FS in the first and third quadrants of the k-space.
In the case of a large FS curvature, on the other hand, the hotspot for the nesting is confined to the vicinity of the gap nodes at ${\bf k}\sim\pm(\pi/2,\pi/2)$.
As a result, in high fields where the nesting hotspot is smaller than the k-space region where a SC order is suppressed by PPB, the overlap of SC and SDW orders in k-space become significantly small, as shown in Fig.\ref{fig:t2FS}(a).

This difference in the k-space structure of the SC and SDW orders due to a FS curvature affects the field dependence of the SDW ordering.
The field dependence of $m$ in the case of the small FS curvature ($t_2/t_1=0.05$) is shown in Fig.\ref{fig:OPvsH}(a), indicating that $m$ is maximal at $H_{c2}$, at which the PPB effect is the most effective.
In this case, the k-space regions contributing to the SDW ordering become wider than the k-space regions in which the SC order is suppressed by PPB, leading to the monotonous increase of $m$ with the field in the SC phase.
In the case of the large FS curvature ($t_2/t_1=-0.05$), on the other hand, the field dependence of $m$ is plotted in Fig.\ref{fig:t2FS}(b), showing that $m$ is maximal not at $H_{c2}$, but at the field \textit{slightly below} $H_{c2}$.
Because the overlap of SC and SDW orders in k-space becomes significantly small at a high field, the SDW ordering due to the mutual enhancement of these orders in k-space is destabilized in high fields. 
This results in the steep decrease of $m$ in high fields and thus the peak of $m$ at the field slightly below $H_{c2}$.

\section{Summary and Discussions} 
\label{sec:summary}

In this paper, we have theoretically analyzed k-space structures of coexisting $d_{x^2-y^2}$-wave SC and SDW orders induced by PPB, and discussed the details of the mechanism of the PPB-induced SDW ordering. 
It has been shown that the SC order is suppressed in the k-space regions near the gap nodes, where the excitation energy $E_{{\bf k},\sigma}$ in a pure SC phase is negative for $\sigma=\uparrow$ due to the Zeeman energy, and that the major contributions to the SDW ordering come from the k-space regions, where the SC order is suppressed by PPB, and the nesting condition of FS ($\xi_{\bf k+Q}\sim-\xi_{\bf k}$) is satisfied (Fig.\ref{fig:map}).
These results reflect the contributions of quasiparticle pockets\cite{kato_2011_kato_2012}, in which $E_{{\bf k},\sigma}<0$, because SDW ordering is mainly induced by quasiparticle states near the quasiparticle pockets connected by the nesting vector ${\bf Q}$.

However, it has been shown that the sign change of the gap function in k-space ($w_{\bf k+Q}=-w_{\bf k}$) is more essential to the realization of the PPB-induced SDW ordering only in a SC phase. 
Figure \ref{fig:OPvsH} shows that the coexistence of SC and SDW orders is stabilized in the high-field region of a $d_{x^2-y^2}$-wave SC phase with the k-space sign change of the gap function $w_{\bf k}$, while these orders are always competitive in the artificial model in which there is no sign change of $w_{\bf k}$ in k-space. In relation to this, it has also been shown in Ref.\onlinecite{hatakeyama_2011} that a $d_{xy}$-wave SC order, which does not satisfy $w_{\bf k+Q}=-w_{\bf k}$ with ${\bf Q}=(\pi,\pi)$, is always competitive to a SDW order with the modulation vector ${\bf Q}$. This fact also indicates that the k-space sign change of the gap function ($w_{\bf k+Q}=-w_{\bf k}$) is a crucial factor for the PPB-induced coexistence of SC and SDW orders.

The importance of the k-space sign change of the gap function can be explained based on the k-space structure of SC and SDW orders.
It has been shown that, in the $d_{x^2-y^2}$-wave SC model, SC and SDW orders are enhanced with each other even in k-space, while, in the artificial model, these orders are competitive in k-space (Fig.\ref{fig:FH}).
This result indicates that this mutual enhancement of $d_{x^2-y^2}$-wave SC and SDW orders originates from the sign change of $w_{\bf k}$ and stabilizes the overlapped structure of these orders in k-space.
As a result, the PPB-induced coexistent phase of SC and SDW orders is largely stabilized by this mutual enhancement.
We stress that this mutual enhancement of SC and SDW orders in k-space resulting from the sign change of the gap function is the dominant mechanism for the PPB-induced SDW ordering.

In the case of a large FS curvature, the hotspot for the nesting becomes smaller than the k-space region where a SC order is suppressed by PPB as the field is  increased, leading to a decrease of the overlap of these orders in k-space (Fig.\ref{fig:t2FS}(a)) and a suppression of SDW ordering in the high-field region of the SC phase.
As a result, the field dependence of a SDW moment $m(H)$ has a peak at a field slightly below $H_{c2}$ (Fig. \ref{fig:t2FS}(b)). 
A peak of SDW ordering at a field slightly below $H_{c2}$ has also been predicted in the case of a moderate strength of PPB with a second order SC transition at a high fields\cite{hatakeyama_2011} and in the case of a strong-coupling SC phase\cite{hatakeyama_2014}. 
However, the peak of SDW ordering in these cases results from the large decrease of a SC order parameter $\Delta$ in higher fields, and the peak of $m(H)$ in the case of a large FS curvature, where $\Delta$ is almost independent of the field (see Fig.\ref{fig:t2FS}(b)), is caused by the different mechanism from them.
That is, the k-space structure of coexisting SC and SDW orders affects the PPB-induced SDW ordering in the present case. Finally, 
We note that, as discussed in Refs.\onlinecite{ikeda_2010a,hatakeyama_2011}, the enhancement of SDW ordering at a field below $H_{c2}$ should be ultimately related to the field-induced antiferromagnetic quantum critical phenomena near $H_{c2}$ observed in several heavy fermion superconductors with a strong PPB effect, such as CeCoIn$_5$\cite{paglione_2003,ronning_2005,singh_2007}, CeRhIn$_5$\cite{park_2010}, Ce$_2$PdIn$_8$\cite{dong_2010,tokiwa_2011}, and NpPd$_5$Al$_2$\cite{honda_2008}.

\begin{acknowledgments}
One of the authors (Y.H.) thanks S. Fujimoto and N. Kawakami for discussions. The research of Y.H. was supported by JSPS Research Fellowship for Young Scientists, and the research of R.I. was supported by Grant-in-Aid for Scientific Research [No.25400368] from MEXT, Japan. 
\end{acknowledgments}


\end{document}